\documentclass[sigconf,natbib=false]{acmart}
\AtBeginDocument{%
  }

\setcopyright{acmlicensed}
\copyrightyear{2026}
\acmYear{2026}
\acmConference[Conference'17]{ACM Conference}{July 2017}{Washington, DC, USA}
\acmDOI{XXXXXXX.XXXXXXX}
\RequirePackage[
  datamodel=acmdatamodel,
  style=acmnumeric,
  sortcites=true,
  ]{biblatex}

\usepackage{algorithm}
\usepackage{graphicx}
\usepackage{algorithmic}
\usepackage{enumitem}

\usepackage{amsmath}
\usepackage{multirow}
\usepackage{listings}
\usepackage[normalem]{ulem}

\lstdefinestyle{instructionoutput}{
  basicstyle=\ttfamily\footnotesize,
  backgroundcolor=\color{black!5},
  frame=single,
  framerule=0pt,
  framesep=4pt,
  breaklines=true,
  breakatwhitespace=true,
  columns=fullflexible,
  keepspaces=true,
  showstringspaces=false,
  aboveskip=4pt,
  belowskip=4pt
}

\begin{document}

\title{LangBP: Language-Guided Reasoning and Acting for Joint Bidding and Pricing}


\settopmatter{authorsperrow=4}

\makeatletter
\newcommand{\deferredauthornote}[1]{%
  \g@addto@macro\@authornotes{%
    \stepcounter{footnote}\footnotetext{#1}}}
\makeatother

\author{Jiaqi Ding}
\authornote{Both authors contributed equally to this research.}
\authornotemark[3]
\email{220252312@seu.edu.cn}
\affiliation{%
  \institution{Southeast University}
  \country{China}
}

\author{Chuan Yang}
\authornotemark[1]
\email{yangchuan56@jd.com}
\affiliation{%
  \institution{JD.com}
  \country{China}
}

\author{Linghui Meng}
\authornote{Corresponding authors.}
\email{menglinghui1@jd.com}
\affiliation{%
  \institution{JD.com}
  \country{China}
}

\author{Shengsheng Niu}
\email{niushengsheng@jd.com}
\affiliation{%
  \institution{JD.com}
  \country{China}
}

\author{Jie He}
\email{hejie67@jd.com}
\affiliation{%
  \institution{JD.com}
  \country{China}
}

\author{Zhangang Lin}
\email{linzhangang@jd.com}
\affiliation{%
  \institution{JD.com}
  \country{China}
}

\author{Ching Law}
\email{lawching@jd.com}
\affiliation{%
  \institution{JD.com}
  \country{China}
}

\author{Xiaolin Fang}
\authornotemark[2]
\email{xiaolin@seu.edu.cn}
\affiliation{%
  \institution{Southeast University}
  \country{China}
}

\deferredauthornote{This work was performed during an internship at JD.com.}

\renewcommand{\shortauthors}{Jiaqi Ding et al.}

\begin{abstract}
Auto-bidding is a long-horizon sequential decision problem for maximizing conversion value under budget and key performance indicator (KPI) constraints.
Recent work extends this task from bidding alone to joint bidding and pricing, where a policy controls bidding decisions and pricing corrections.
Existing methods mainly rely on numerical trajectory modeling, which offers limited support for interpreting campaign context and expressing high-level strategies. Large language models (LLMs) can complement this paradigm with their reasoning capabilities.
However, existing language-guided methods have two limitations.
First, they condition actions on language strategies without modeling the corresponding state changes, making it difficult to distinguish errors in strategy understanding from errors in action generation.
Second, different instructions can produce similar execution effects, leading to imbalanced policy updates across effects.
We propose \textbf{LangBP}, a hierarchical framework for language-guided joint bidding and pricing.
LangBP's Semantic Decision Transformer (\textbf{S-DT}) predicts target states from the instruction and the trajectory history, then recovers the joint action via inverse dynamics.
We further propose Execution-Grouped Policy Optimization (\textbf{EGPO}), which scores candidate effects with a Context--Effect Verifier (\textbf{CEV}) and balances policy updates across effect groups.
Experiments on AuctionNet show that LangBP outperforms strong baselines, and online A/B tests further demonstrate business gains in real-world deployment on a large-scale e-commerce platform.
\end{abstract}




\maketitle

\begin{figure}[H]
\centering
\includegraphics[width=\columnwidth]{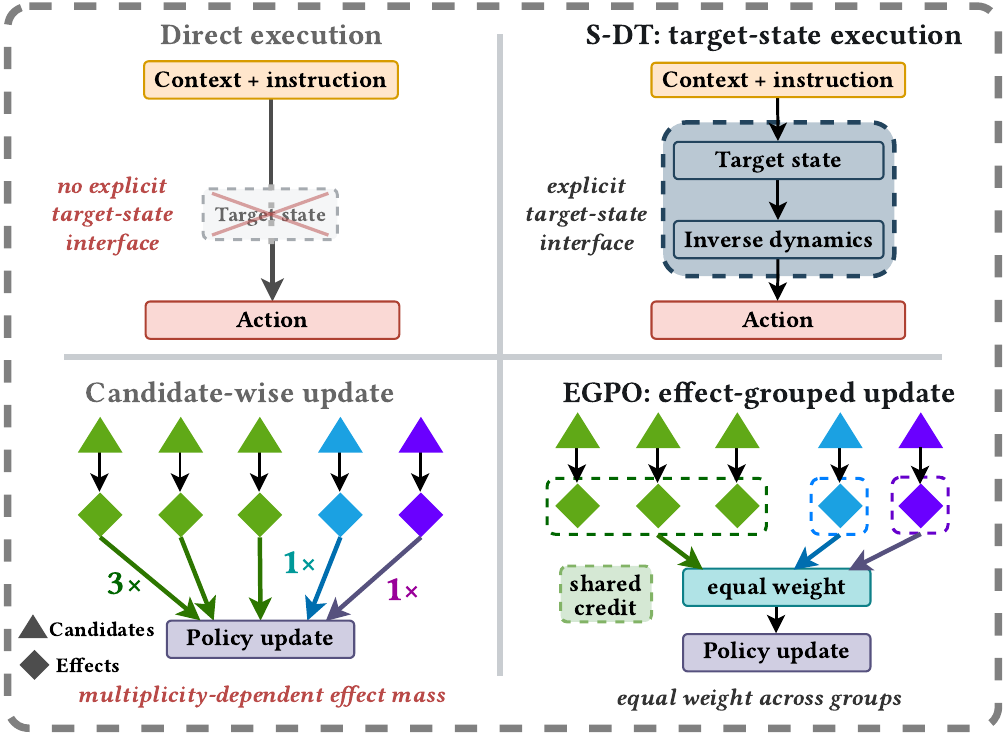}
\caption{Design of LangBP. Left: direct execution and candidate-wise updates; right: S-DT target-state execution and EGPO effect-grouped updates.}
\Description{The top row contrasts direct action execution, where the target state is not on the action path, with S-DT target-state planning followed by inverse dynamics. The bottom row shows three candidate instructions producing one effect and one instruction producing another. Candidate-wise optimization assigns multiplicity-dependent effect mass, whereas EGPO groups similar effects and assigns equal outer mass across active groups.}
\label{fig:intro_motivation}
\end{figure}

\section{Introduction}
\label{sec:introduction}

Auto-bidding is a core technology in online advertising.
Advertisers only need to specify a campaign budget and key performance indicator (KPI) constraints at launch.
An auto-bidding policy then adjusts bidding parameters to maximize conversion value while satisfying these constraints~\cite{cai2017rtb,he2021uscb}.
Each decision affects future spending, constraint status, and available traffic~\cite{mou2022sorl}.
Therefore, auto-bidding is a long-horizon sequential decision problem~\cite{su2024auctionnet}.
Recent methods use reinforcement learning or generative sequence models to learn auto-bidding policies from logged trajectories~\cite{wu2018budget,guo2024diffbid,gao2025gave}.
Decision Transformer (DT)~\cite{chen2021decision} generates actions conditioned on returns-to-go, past states, and past actions.
It is a representative method for offline continuous control.

Earlier work studies hybrid auction formats and performance-based pricing models in online advertising~\cite{zhu2011hybrid,hu2016pricing}.
Recent work extends bidding-only control to joint bidding and pricing.
The policy jointly controls bidding decisions and subsequent pricing corrections~\cite{meng2026jdbp}.
Bids affect access to future traffic, while pricing corrections reduce accumulated constraint deviations.
A generative framework can coordinate bidding and pricing to pursue future value and correct past deviations. This avoids relying on bidding alone to balance both goals.

Existing joint bidding--pricing work remains based on numerical trajectory modeling~\cite{meng2026jdbp}. This paradigm has been widely used for continuous decision-making~\cite{chen2021decision,janner2021trajectory,zhang2026qga}, but does not explicitly reason about the campaign context behind the trajectories.
Large language models (LLMs) can instead analyze campaign contexts, summarize feedback, and express high-level strategies~\cite{li2026lbm,zhu2026sembid,lv2026decisionllm}.
Recently, LBM~\cite{li2026lbm} separates auto-bidding into high-level reasoning and low-level execution.
LBM-Think first summarizes historical bidding status and reasons about an adjustment direction.
LBM-Act then generates continuous actions from the reasoning output and state information.
SemBid~\cite{zhu2026sembid} takes a different approach. It integrates semantic tokens for task, history, and strategy with bidding trajectories in a DT.
Both methods use language to guide continuous action generation.

However, one challenge is to determine the state change corresponding to a language strategy.
Existing methods use language to provide high-level strategy guidance, but do not explicitly represent the corresponding state changes~\cite{li2026lbm,zhu2026sembid}. They directly predict continuous actions from language information and numerical trajectories and supervise these predictions with logged actions.
Such action-level supervision provides no direct learning signal for these state changes~\cite{liu2022depo}.
Thus, the loss does not explicitly distinguish errors in strategy understanding from those in action generation.

Several methods first predict states and then generate actions, but do not incorporate language guidance into state prediction.
DePO~\cite{liu2022depo} predicts a target state and recovers the corresponding action through inverse dynamics.
In auto-bidding, DiffBid~\cite{guo2024diffbid} generates bidding actions from predicted future states through inverse dynamics. GUIDE~\cite{zhang2026guide} uses a similar design to produce behavior-consistent fallback actions.

A second challenge is to provide execution feedback to a language policy.
LBM~\cite{li2026lbm} scores each candidate reasoning output using the relative-Q between its induced action and the logged action.
It then updates the language policy with the candidate that has the highest positive relative-Q.
This provides offline execution feedback, but uses only one candidate per update despite multiple effective strategies for a context.
Group Relative Policy Optimization (GRPO)~\cite{shao2024deepseekmath} learns from multiple candidates sampled for the same context. Later work extends this idea to sequential language agents~\cite{feng2025gigpo}.
Each candidate instruction induces an execution effect, defined by its joint action and predicted target-state change. Different instructions may produce similar execution effects.
When these instructions are optimized separately, similar execution effects can influence the policy update multiple times.
Consequently, execution effects induced by more candidate instructions receive more total optimization weight within the sampled set, which can weaken learning signals for other valuable effects~\cite{anschel2025gapo,sinha2026modecollapse}.
The left panels of Figure~\ref{fig:intro_motivation} illustrate these two challenges.

To translate language strategies into continuous control, we propose \textbf{LangBP} (\textbf{Lang}uage-Guided Reasoning and Acting for Joint \textbf{B}idding and \textbf{P}ricing).
LangBP contains a high-level instruction policy and a low-level continuous executor.
The instruction policy generates a structured instruction from the campaign context, containing a strategy explanation, bidding direction, and pricing strength.
At the lower level, the Semantic Decision Transformer (\textbf{S-DT}) connects language instructions to continuous actions through target states. Target states represent the next bidding and pricing states. S-DT predicts these states from the instruction and the numerical trajectory history. Inverse dynamics then recovers the corresponding joint bidding--pricing action. The instruction affects the joint action only through these target states, thereby structurally separating target-state planning from action generation.

To provide effect-balanced execution feedback to the instruction policy, we propose Execution-Grouped Policy Optimization (\textbf{EGPO}). For each campaign context, the instruction policy samples multiple candidates, which a frozen S-DT maps to execution effects. S-DT also produces a base execution effect from a fixed blank instruction.
Offline trajectories do not directly reveal the quality of newly sampled execution effects. We therefore introduce the Context--Effect Verifier (\textbf{CEV}), which learns context-dependent scores from pairs constructed from locally matched, behavior-supported trajectory outcomes.
EGPO groups candidates with similar execution effects and computes each group's advantage by comparing its candidate effects with the base effect using CEV. It shares credit within groups and balances updates across groups. This reduces the extra optimization weight assigned to effects induced by more candidates and preserves learning signals for valuable alternatives.
The right panels of Figure~\ref{fig:intro_motivation} summarize the corresponding S-DT and EGPO designs.

Our main contributions are as follows:
\begin{itemize}[
    leftmargin=*,
    topsep=3pt
]
    \item To translate language strategies into continuous control, we propose LangBP, a hierarchical framework with a high-level instruction policy and a low-level continuous executor.

    \item We design S-DT to execute high-level language instructions by predicting their target states. It then uses inverse dynamics to recover the corresponding joint bidding--pricing action. By modeling state changes, this formulation allows language instructions to guide continuous control through explicit target states.

    \item We introduce EGPO to provide execution feedback to the instruction policy. 
    It groups candidates by execution effect and balances policy updates across groups, while CEV learns context-dependent execution preferences from offline trajectories. This balances optimization mass across distinct effects per sampled set and preserves learning signals for valuable alternatives.

    \item Benchmark experiments compare LangBP with existing methods and analyze each component.
    Online A/B tests in a large-scale e-commerce advertising system show deployment gains.
\end{itemize}

\section{Related Work}

\subsection{Auto-Bidding and Joint Bidding--Pricing}
Auto-bidding optimizes long-horizon conversion value under budget and KPI constraints.
Early reinforcement learning methods learn numerical policies for constrained bidding, including unified constraint control and safe online refinement~\cite{wu2018budget,he2021uscb,mou2022sorl}.
Generative decision models instead learn from logged trajectories.
Decision Transformer provides a representative sequence-modeling backbone~\cite{chen2021decision}.
DiffBid generates future state trajectories with conditional diffusion and recovers bids through inverse dynamics~\cite{guo2024diffbid}, while GAS, GRAD, GAVE, and QGA improve policy search or exploration through post-training search, structured action exploration, and value guidance~\cite{li2025gas,lei2026grad,gao2025gave,zhang2026qga}.
These methods differ in trajectory modeling and policy improvement but remain bidding-only.

Allocation and payment can also be coupled.
Hybrid auctions and performance-based payment models link allocation and charging rules at the auction level~\cite{zhu2011hybrid,hu2016pricing}.
JD-BP introduces campaign-level post-auction pricing correction, separating future traffic acquisition from the repair of accumulated KPI deviations~\cite{meng2026jdbp}.
LangBP builds on this joint action space and studies how language-guided reasoning coordinates bidding and pricing.

\subsection{Language-Guided Execution Optimization}
Recent methods connect language to continuous control through semantic conditioning or hierarchical execution.
SemBid inserts task, history, and strategy semantics into bidding trajectories, while DecisionLLM conditions action generation on task descriptions and numerical histories~\cite{zhu2026sembid,lv2026decisionllm}.
SAGE aligns textual strategies with bidding trajectories~\cite{cai2026sage}, whereas LBM and AIGB-R1 separate high-level reasoning or planning from low-level numerical execution~\cite{li2026lbm,dou2026aigbr1}.
A complementary line uses predicted states as an intermediate control representation.
DePO recovers actions from predicted target states~\cite{liu2022depo}; in auto-bidding, DiffBid and GUIDE combine future-state prediction with inverse dynamics~\cite{guo2024diffbid,zhang2026guide}.
LangBP combines these directions by using language-conditioned target states as the interface to joint bidding--pricing actions.

Language-policy optimization also requires feedback from numerical execution.
LBM ranks reasoning candidates with an offline Q-function~\cite{li2026lbm}.
AIGB-R1 uses simulator feedback for planner--executor optimization, while AIGB-Pearl learns a trajectory evaluator for generative policy search~\cite{dou2026aigbr1,mou2026pearl}.
GRPO provides relative feedback across sampled responses, and GiGPO extends this principle to sequential language agents~\cite{shao2024deepseekmath,feng2025gigpo}.
When multiple text candidates induce similar execution effects, candidate-level objectives can assign more total update weight to more frequently represented modes~\cite{anschel2025gapo,sinha2026modecollapse}.
LangBP instead groups candidates by execution effect, defined by the joint action and predicted target-state change, and balances policy updates across effect groups.

\section{Problem Formulation}
\label{sec:problem}
Section~\ref{sec:jointrecap} defines joint bidding and pricing under full knowledge of future auction opportunities.
Section~\ref{sec:sequential_control} casts this formulation as a language-guided sequential control problem when future opportunities are unknown.

\subsection{Joint Bidding and Pricing Correction}
\label{sec:jointrecap}

\paragraph{Bidding-only formulation and its limitation.}
Given $N$ impression opportunities arriving sequentially, the classic auto-bidding problem chooses a winning indicator $x_i\in\{0,1\}$ for each opportunity $i$ to maximize accrued value subject to a budget constraint and a set of KPI constraints~\cite{meng2026jdbp}:
\begin{equation}\label{eq:oripro}
\begin{aligned}
\max_{x_i}\ &\textstyle\sum_i x_i v_i\\
\text{s.t.}\ &\textstyle\sum_i x_i c_i\le B,\quad x_i\in\{0,1\},\ \forall i,\\
&\textstyle\sum_i x_i c_i\le\rho_j\textstyle\sum_i x_i\gamma_{ij},\ \forall j,
\end{aligned}
\end{equation}
where $v_i$ is the estimated value of opportunity $i$, $c_i$ is the payment charged by the auction mechanism, $B$ is the total budget, $\gamma_{ij}$ is the $j$-th KPI outcome of opportunity $i$, and $\rho_j$ is its target bound. Under perfect knowledge of future values and a stationary market, the optimal policy admits a closed-form linear bid,
\begin{equation}\label{eq:optbid0}
bid_i=\lambda_0v_i+\textstyle\sum_j\rho_j\lambda_j\gamma_{ij},
\end{equation}
with $\lambda_0,\lambda_j$ the dual variables of the budget and KPI constraints~\cite{meng2026jdbp}. In practice, however, prediction error, conversion delay, and non-stationary competition make constraint violations unavoidable. Because \eqref{eq:oripro} exposes only one control lever, $x_i$, a bidding-only policy can repay an accrued violation only by bidding more conservatively. Future value maximization and the repair of past deviations must therefore compete for the same variable.

\paragraph{Joint bidding--pricing.}
A growing line of work resolves this coupling by separating the repair of past deviations from the pursuit of future value through an auxiliary, post-auction lever.
The need for such a correction is not specific to auto-bidding: classical dynamic mechanism design shows that incentive-compatibility and individual-rationality constraints need only hold \emph{on average} over a sequence of transactions rather than for every single transaction~\cite{mirrokni2016dynamic}. Concretely, the authors augment a sequence of otherwise-independent, single-shot auctions with a \emph{bank account} balance $bal_t$ that accumulates the buyer's historical utility deficits or surpluses, and let payment rule depend on this running balance rather than on the current transaction alone, with
\begin{equation}\label{eq:bankupdate}
bal_{t+1}=bal_t-\operatorname{spend}_t(bal_t)
+\operatorname{deposit}_t(bal_t,v_t),
\end{equation}
where $\operatorname{spend}_t$ and $\operatorname{deposit}_t$ govern how the bank-account balance is drawn down or replenished. They prove that any mechanism restricted to price each transaction independently, i.e.\ with $bal_t\equiv0$ and no such history-dependent correction, is provably dominated in revenue and social welfare by mechanisms that permit this bank-account correction. This result formalizes, in a general auction-theoretic setting, the same intuition that motivates our use of a payment correction: a payment rule that reacts only to the current transaction is structurally suboptimal once constraint satisfaction is evaluated over a horizon rather than per query, and a correction term evaluated against the accumulated history is required to close this gap.

JD-BP~\cite{meng2026jdbp} is among the most effective instantiations of this idea in the auto-bidding setting: it plays a role analogous to the spend policy $\operatorname{spend}_t$ above, but augments the bid with a bounded, non-positive \emph{pricing correction} applied after the auction has settled, drawing down an explicitly tracked historical deficit $bal$ (Eq.~\ref{eq:baldef}) through the payment channel rather than through the bid alone. Let $t=t_m,\dots,T$ index the opportunities remaining after the current decision step $t_m$, with $x_t,v_t,c_t$ retaining the meanings above. Let $B_{t_m}$ denote the remaining budget. The pricing correction $y_t\le0$ turns the auction-determined payment $c_t$ into an actual charge $p_t=c_t+y_t$. The elapsed history already fixes a \emph{historical deficit}
\begin{equation}\label{eq:baldef}
bal=\max\Big(0,\ \textstyle\sum_{t<t_m}x_t c_t-\rho\textstyle\sum_{t<t_m}x_t v_t\Big)\ge0,
\end{equation}
and bidding and pricing are jointly optimized as
\begin{equation}\label{eq:jom}
\begin{aligned}
\max_{x_t,y_t}\ &\textstyle\sum_{t=t_m}^{T}x_t v_t\\
\text{s.t.}\ &\textstyle\sum_{t=t_m}^{T}x_t(c_t+y_t)\le B_{t_m},\quad
\textstyle\sum_{t=t_m}^{T}x_t c_t\le\rho\textstyle\sum_{t=t_m}^{T}x_t v_t,\\
&\textstyle\sum_{t=t_m}^{T}x_t y_t+bal=0,\quad x_t\in\{0,1\},\ y_t\le0.
\end{aligned}
\end{equation}
JD-BP shows that \eqref{eq:jom} is equivalent to a bidding-only problem with a deficit-enlarged budget $B_{t_m}+bal$, so the optimal bid retains the closed form of \eqref{eq:optbid0},
\begin{equation}\label{eq:optbidjoint}
bid_t=\hat\lambda_0v_t+\textstyle\sum_j\rho_j\hat\lambda_j\gamma_{tj},
\end{equation}
which, specialized to the single ROI ratio $\rho$ used in \eqref{eq:jom}, reduces to the value-proportional bid $bid_t=(\hat\lambda_0+\rho\hat\lambda_\rho)v_t$; any refund schedule that repays the deficit exactly is a feasible pricing correction, e.g.\ the uniform rule $y_t=-bal/\sum_{\tau=t_m}^{T}x_\tau$.
The resulting joint value weakly dominates the bidding-only value, $V_O^\star\le V_J^\star$~\cite{meng2026jdbp}. Routing the historical deficit through pricing never hurts and strictly helps once a deficit has accrued. We therefore adopt JD-BP's joint action space, $a_t=(a_t^{\mathrm{bid}},a_t^{\mathrm{price}})$, where $a_t^{\mathrm{bid}}$ is the continuous bidding parameter applied in the current interval and $a_t^{\mathrm{price}}\le0$ requests a post-auction pricing correction. After settlement, the request is projected onto the feasible charge range to realize $y_t\le0$ and $p_t=c_t+y_t$. The bidding adjustment is the change in $a_t^{\mathrm{bid}}$ from its preceding value. This joint action provides the executable interface for LangBP.

\subsection{Language-Guided Sequential Control}
\label{sec:sequential_control}

The closed-form solution in Eq.~\eqref{eq:optbidjoint} assumes exact future opportunity values and a stationary market.
Because traffic, competition, and value estimates change during delivery, the joint problem must instead be solved from observed campaign history~\cite{su2024auctionnet,meng2026jdbp}.
At decision step $t$, state $s_t$ summarizes delivery history, remaining budget, and constraint status.
The feasible joint action $a_t=(a_t^{\mathrm{bid}},a_t^{\mathrm{price}})$ uses the controls in Section~\ref{sec:jointrecap}: bidding affects auctions in the current interval, while pricing is fixed from the same pre-interval information and applied after settlement.
The environment then emits an interval reward $r_t$ and transitions to $s_{t+1}$.

$r_t$ is the interval conversion value with hinge penalties on violations of $B/T$ and the payment/KPI bounds in Eq.~\eqref{eq:jom}. These penalties shape training but preserve the constrained campaign objective.

Let $H_t=(s_{t-h:t},a_{t-h:t-1},r_{t-h:t-1})$ be the pre-decision history over the previous $h$ transitions.
Numerical pretraining uses the hindsight label $\hat R_t=\sum_{k=0}^{T-t}\eta^k r_{t+k}$, with discount $\eta\in(0,1]$.
Because it contains future outcomes, $\hat R_t$ is confined to numerical pretraining and excluded from semantic alignment, policy optimization, the instruction-policy and teacher inputs, and CEV.
Later stages use $R_t$, initialized from a validation-selected training-return percentile and fixed before delivery, then updated causally as $R_{t+1}=(R_t-r_t)/\eta$.
This recurrence uses only the initial target and previously observed rewards, never the logged future suffix.
The next state is sampled from unknown auction dynamics conditioned on $(H_t,a_t)$.

Let context $\chi_t$ collect the advertiser objective, KPI targets, budget plan, and history $H_t$ known before decision $t$; the instruction policy samples $z_t\sim\pi_\theta(\cdot\mid\chi_t)$.
Here $H_t$ is the rolling numerical state--action--reward history, whereas $\chi_t$ augments it with campaign-level objectives and planning information for language generation and context matching.
After numerical pretraining, the executor maps $(H_t,R_t,z_t)$ to a feasible joint action through an explicit target-state prediction.
The hierarchy is trained from offline trajectories while future outcomes remain outside the instruction-policy and CEV inputs.
Section~\ref{sec:sdt} defines the executor, and Sections~\ref{sec:cev}--\ref{sec:egpo} define offline execution feedback.

\section{Methodology}
\begin{figure*}[!t]
  \centering
  \includegraphics[width=\textwidth]{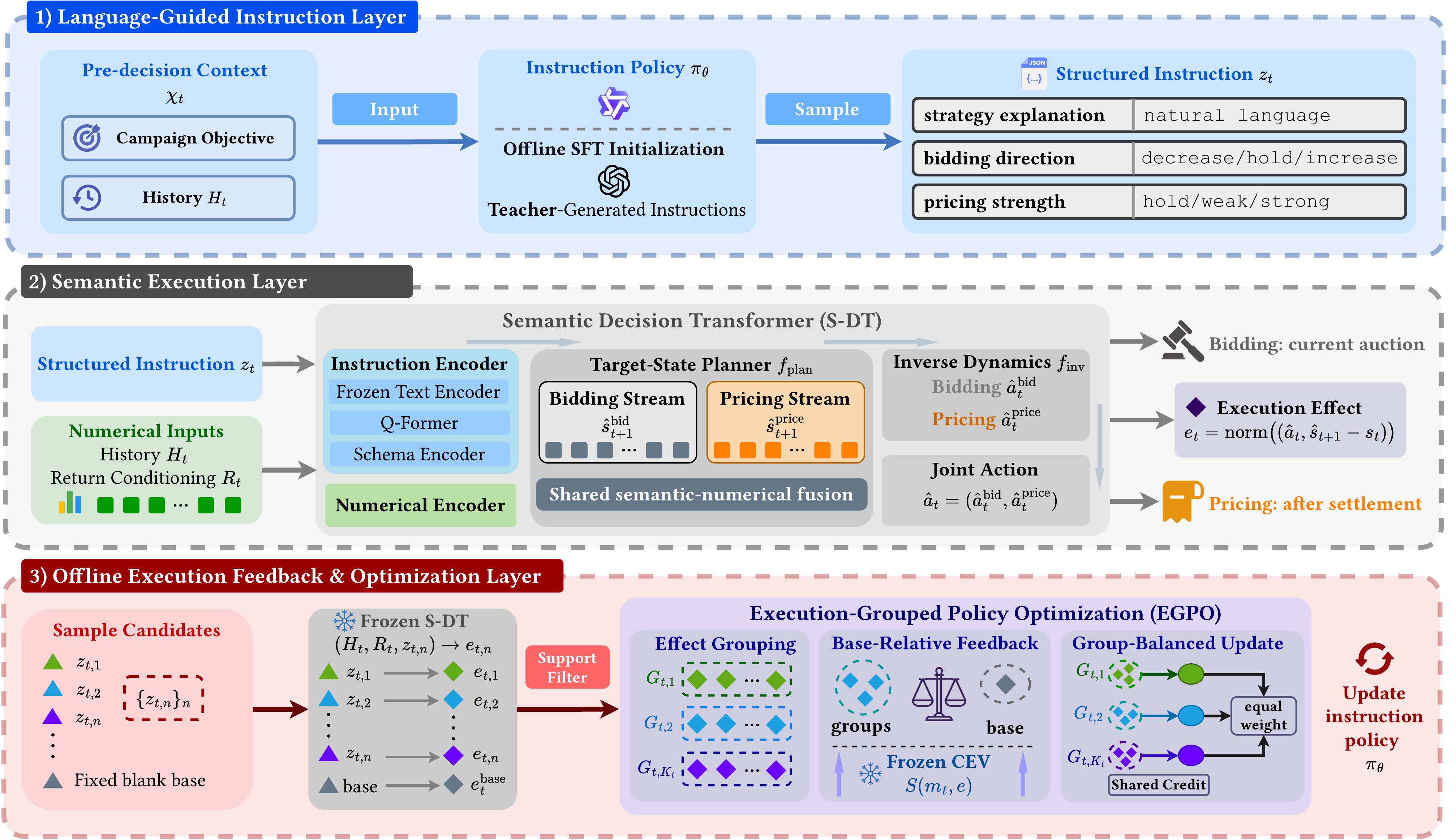}
  \caption{Overview of LangBP. S-DT executes instructions, while CEV-guided EGPO updates the policy over effect groups.}
  \Description{A three-layer framework diagram. The instruction layer maps campaign context to a structured instruction. The semantic execution layer uses S-DT to predict bidding and pricing target states, recover a joint action, and construct a normalized execution effect. The offline feedback layer applies local support filtering, CEV scoring, and EGPO grouping to update only the instruction policy.}
\label{fig:training_pipeline}
\end{figure*}

\begin{figure*}[!t]
\centering
\includegraphics[width=\textwidth]{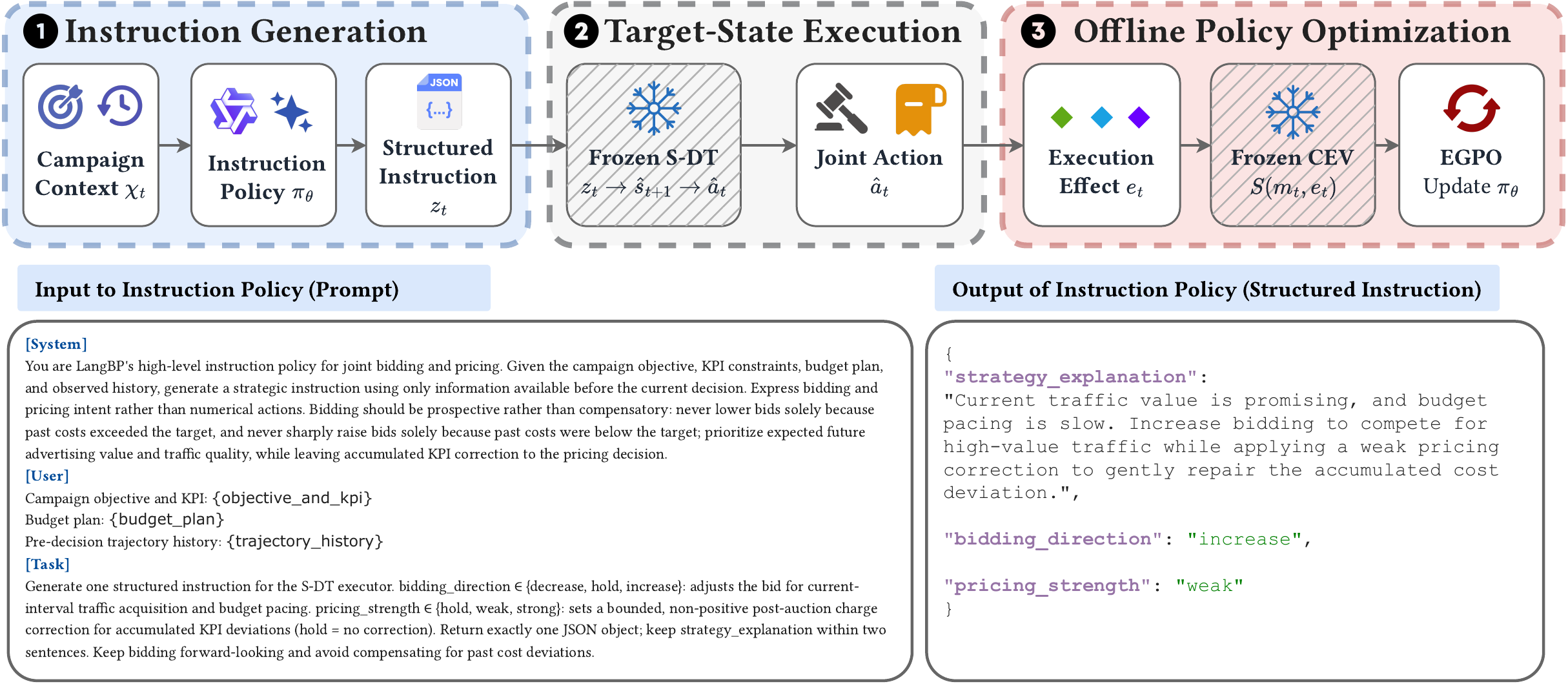}
\caption{End-to-end workflow of LangBP.}
\Description{The campaign context is converted into a structured instruction containing a strategy explanation, bidding direction, and pricing strength. S-DT executes the instruction as a joint bidding and pricing action, CEV evaluates the resulting execution effect, and EGPO feeds the evaluation back to update the instruction policy. Prompt and structured-output examples are shown below the workflow.}
\label{fig:end_to_end_workflow}
\vspace{-4pt}
\end{figure*}

Figure~\ref{fig:training_pipeline} overviews LangBP's architecture, while Figure~\ref{fig:end_to_end_workflow} shows the end-to-end workflow, instruction prompt, and structured output.

\subsection{Policy Initialization}
\label{sec:policy_init}
From each logged transition, we categorize the bidding change $a_t^{\mathrm{bid}}-a_{t-1}^{\mathrm{bid}}$ as \texttt{decrease}, \texttt{hold}, or \texttt{increase} using symmetric thresholds derived from a training-split quantile, and categorize the pricing magnitude $|a_t^{\mathrm{price}}|$ as \texttt{hold}, \texttt{weak}, or \texttt{strong} using two ordered quantile thresholds.
All thresholds are fixed from the training split.

A frozen teacher LLM receives only the pre-decision context $\chi_t$ and the two labels and generates a concise strategy explanation grounded in $\chi_t$.
We retain only schema-valid outputs that match both control labels and pass a deterministic context-consistency check.
Neither generation nor filtering uses $\hat R_t$ or later outcomes.

The retained structured instruction $z_t$, containing a strategy explanation, a bidding-direction label, and a pricing-strength label, supervises a smaller instruction policy~\cite{wang2023selfinstruct}.
The categorical fields encode coarse controls, while the explanation field of $z_t$ distinguishes context-dependent strategies sharing the same labels; S-DT consumes both.
The teacher is used only to construct offline data; the student supports deployment and repeated EGPO sampling.

S-DT and CEV are trained from logged trajectories, then frozen while EGPO updates only the instruction policy.
At deployment, the instruction policy and S-DT produce one joint action per decision.

\subsection{Semantic Decision Transformer}
\label{sec:sdt}
S-DT follows a two-stage path: the instruction and numerical history predict a target state, from which inverse dynamics recovers the joint action (Figure~\ref{fig:training_pipeline}).
Because the inverse model never reads $z_t$, language affects the action only through this target state.
A frozen text encoder and Q-Former compress the explanation~\cite{li2023blip2}; a schema encoder represents its controls, and a numerical encoder represents $(H_t,R_t)$ after pretraining on $(H_t,\hat R_t)$.
Their fusion enters a dual-stream planner~\cite{liu2022depo,zhang2026guide}: bidding and pricing views summarize traffic acquisition and budget pacing, and charge correction and KPI deviations, respectively~\cite{meng2026jdbp}, while exchanging shared time and budget information.

The planner $f_{\mathrm{plan}}$ and inverse model $f_{\mathrm{inv}}$ implement
\begin{equation}
\label{eq:sdt_mapping}
\begin{aligned}
\hat s_{t+1}
&=f_{\mathrm{plan}}(H_t,R_t,z_t),\\
\hat a_t
&=f_{\mathrm{inv}}(H_t,\hat s_{t+1}).
\end{aligned}
\end{equation}
Here $\hat s_{t+1}$ contains bidding and pricing views; two inverse heads produce $\hat a_t=(\hat a_t^{\mathrm{bid}},\hat a_t^{\mathrm{price}})$, projected to the feasible ranges in Section~\ref{sec:jointrecap}, without assuming invertible auction dynamics.

For instruction $z_t$, define the normalized execution effect
\begin{equation}
\label{eq:sdt_effect}
e_t(z_t)=\operatorname{norm}\!\left(
\bigl(\hat a_t(z_t),\hat s_{t+1}(z_t)-s_t\bigr)
\right).
\end{equation}
Here $\operatorname{norm}(\cdot)$ uses training-set component statistics; the resulting effect $e_t(z_t)$ records the predicted joint action and expected state displacement for CEV and EGPO.

\paragraph{Training objective.}
Let $I_t^{\mathrm{bid}}$ and $I_t^{\mathrm{price}}$ denote the feasible bidding-adjustment and pricing-magnitude intervals implied by the current instruction labels.
Instruction consistency is
\begin{equation}
\label{eq:sdt_inst}
\mathcal L_{\mathrm{inst}}
=\operatorname{dist}(\hat a_t^{\mathrm{bid}}-a_{t-1}^{\mathrm{bid}},I_t^{\mathrm{bid}})^2
+\operatorname{dist}(|\hat a_t^{\mathrm{price}}|,I_t^{\mathrm{price}})^2,
\end{equation}
where $\operatorname{dist}$ is zero inside an interval and otherwise measures distance to its nearest endpoint.
With frozen, differentiable $f_{\mathrm{inv}}$, this loss $\mathcal L_{\mathrm{inst}}$ reaches $f_{\mathrm{plan}}$ only via $\hat s_{t+1}$, preserving the two-stage path.

For observed transitions, we use three complementary objectives for target-state prediction, inverse-action recovery, and dynamics consistency.
For an observed transition $(H_t,a_t,s_{t+1})$, define
\begin{equation}
\label{eq:sdt_supervised_losses}
\begin{aligned}
\mathcal L_{\mathrm{state}}&=\|\hat s_{t+1}-s_{t+1}\|_2^2,\\
\mathcal L_{\mathrm{action}}&=
\|f_{\mathrm{inv}}(H_t,s_{t+1})-a_t\|_2^2
+\alpha\|f_{\mathrm{inv}}(H_t,\hat s_{t+1})-a_t\|_2^2,\\
\mathcal L_{\mathrm{dyn}}&=\|f_{\mathrm{dyn}}(H_t,\hat a_t)-\hat s_{t+1}\|_2^2.
\end{aligned}
\end{equation}
Here, $f_{\mathrm{inv}}$ recovers actions from history and next-state targets, while $f_{\mathrm{dyn}}$ predicts next states from history and actions; both are learned from logs and frozen before semantic alignment.
$\mathcal L_{\mathrm{state}}$ anchors planned targets to observed next states without claiming ground-truth intent or causal targets.
The two $\mathcal L_{\mathrm{action}}$ terms train recovery from observed next states and require planner targets to recover the same logged action, weighted by $\alpha\geq0$.
$\mathcal L_{\mathrm{dyn}}$ checks whether the recovered action $\hat a_t$ realizes the planned target $\hat s_{t+1}$ under $f_{\mathrm{dyn}}$.

Counterfactual instruction examples replace one or both labels, after which the teacher regenerates the explanation field of $z_t$ from the same pre-decision context.
We retain only examples whose predicted actions remain within validation-selected local support of logged behavior.
Because their counterfactual next states are unobserved, they receive instruction and dynamics supervision but no state or action target.

\paragraph{Blank-instruction base and training stages.}
The fixed blank instruction uses an empty explanation and \texttt{no-guidance} controls to define EGPO's base normalized effect $e_t^{\mathrm{base}}$.
Observed, counterfactual, and blank examples use the applicable losses above.
Numerical pretraining learns $f_{\mathrm{plan}}$'s numerical backbone, $f_{\mathrm{inv}}$, and $f_{\mathrm{dyn}}$ from logged trajectories using $\hat R_t$ without language; semantic alignment switches to $R_t$, freezes them, and trains only the instruction-conditioned planner paths, Q-Former, and schema encoder.

\subsection{Context--Effect Verifier}
\label{sec:cev}
Candidate instructions have predicted effects but no observed outcomes.
CEV therefore learns observational preferences in three steps: local-support filtering, matched historical pairs, and context-conditioned ranking.

For logged transition $i$, the observed normalized effect is $e_i=\operatorname{norm}\bigl((a_i,s_{i+1}-s_i)\bigr)$, where $\operatorname{norm}$ standardizes effect coordinates; frozen S-DT predicts $e_t(z_t)$ for candidate $z_t$.
Equation~\eqref{eq:sdt_supervised_losses} aligns observed and predicted coordinates, so logged-effect orderings within local support serve only as observational ranking signals.

\paragraph{Local-support filtering.}
The matching vector $m$, a standardized representation of context $\chi$, summarizes campaign objectives, budget/time coordinates, and recent trajectory history.
CEV first restricts the training log to transitions at the same decision step whose matching vectors are close to $m$, and then finds the $K$ logged effects closest to query effect $e$ within this context-local pool.
We denote these neighbors by $\mathcal N_K(e\mid t,\chi)$, fix $K$ on validation, and for effect vectors $x,y$ use dimension-normalized RMS, $\operatorname{RMS}(x,y)=\|x-y\|_2/\sqrt{\dim(x)}$, where $\dim(x)$ is their number of coordinates.
The local effect-support distance is
\begin{equation}
\label{eq:cev_support}
D_{\mathrm{sup}}(e\mid t,\chi)=\frac{1}{K}
\sum_{i\in\mathcal N_K(e\mid t,\chi)}
\operatorname{RMS}(e,e_i).
\end{equation}
A normalized effect $e$ is retained only when $\mathcal N_K(e\mid t,\chi)$ contains $K$ neighbors and $D_{\mathrm{sup}}(e\mid t,\chi)$ is below a fixed cutoff; this prevents CEV and EGPO from extrapolating to effects unsupported by logged behavior under similar campaign conditions~\cite{sun2022knn,fujimoto2019bcq,zhan2024offline}.

CEV uses separate MLP towers and maps their embeddings and product to scalar observational score $S(m,e)$ as $[\mathrm{MLP}_m(m);\mathrm{MLP}_e(e);\mathrm{MLP}_m(m)\odot\mathrm{MLP}_e(e)]\mapsto S(m,e)$~\cite{he2017neural}.
Here semicolons denote concatenation and $\odot$ element-wise multiplication, which makes the score context dependent while preserving separate representations.

\paragraph{Matched historical pairs.}
Advertisers are split into five folds. For each transition $i$, a residualizer trained on the other four folds predicts the context-predictable discounted suffix utility $\hat U_i$ from pre-action matching vector $m_i$~\cite{chernozhukov2018dml}.
For suffix horizon $L$, we define residualized utility
$U_i=\sum_{k=0}^{L-1}\eta^k r_{i+k}-\hat U_i$, where $\eta$ is the discount factor; $U_i$ measures whether the logged suffix performs above or below its context-predictable level.
Historical pairs $(i,j)$ are formed from distinct logged periods of the same advertiser at the same decision step and budget setting, with similar $m_i$ and $m_j$.
We retain $(i,j)$ only when $e_i$ and $e_j$ are supported under both contexts, sufficiently different, and have a reliable residual-utility gap $|U_i-U_j|$.
If $U_i>U_j$, the ordered pair treats $e_i$ as preferred to $e_j$, and vice versa; the resulting pairs form $\mathcal P$, while contextually closer pairs receive larger weight $w_{ij}$.
Appendix~\ref{sec:appendix_impl} defines the validation-fixed neighborhood sizes and pair-selection thresholds and reports their selected values.

\paragraph{Context--effect ranking.}
To avoid tying a pairwise preference to either matched context alone, we score $e_i$ and $e_j$ under both $m_i$ and $m_j$ and average the two differences.
For each ordered pair $(i,j)\in\mathcal P$, the symmetric score gap $\Delta S_{ij}$ and CEV loss $\mathcal L_{\mathrm{CEV}}$ are
\begin{equation}
\label{eq:cev_objective}
\begin{aligned}
\Delta S_{ij}&=\frac12\bigl[
S(m_i,e_i)-S(m_i,e_j)
+S(m_j,e_i)-S(m_j,e_j)
\bigr],\\
\mathcal L_{\mathrm{CEV}}
&=-\mathbb E_{(i,j)\sim\mathcal P}
\left[w_{ij}\log\sigma(\Delta S_{ij})\right].
\end{aligned}
\end{equation}
This weighted RankNet objective~\cite{burges2005ranknet} scores both effects under both contexts; $\sigma$ is the sigmoid and $\mathbb E$ averages over pairs from $\mathcal P$.
Local matching reduces observed context mismatch but not unobserved confounding; suffix outcomes appear only in pair labels, so $S$ remains an observational, non-causal ranking score.

\subsection{Execution-Grouped Optimization}
\label{sec:egpo}
GRPO learns from several outputs sampled for one context~\cite{shao2024deepseekmath}.
EGPO changes their within-context weighting: similar supported effects share group credit, and active groups receive equal outer weight.
This prevents group multiplicity from increasing outer optimization weight within a sampled set, without changing effect frequencies across contexts.

\paragraph{Grouping.}
For context $\chi_t$, the pre-update instruction policy samples multiple instructions; frozen S-DT maps each $(H_t,R_t,z_{t,n})$ to normalized effect $e_{t,n}$, and Eq.~\eqref{eq:cev_support} filters unsupported effects.
Given a supported base effect $e_t^{\mathrm{base}}$ and at least two retained candidate effects $e_{t,n}$, complete-linkage clustering forms $K_t$ groups $G_{t,1},\ldots,G_{t,K_t}$ whose within-group distances do not exceed validation-selected $\epsilon_g$.

\paragraph{Group feedback.}
Frozen CEV scores every retained candidate effect $e_{t,n}$ and the base effect $e_t^{\mathrm{base}}$ under $m_t$, the matching vector for context $\chi_t$:
\begin{equation}
\label{eq:egpo_group_advantage}
\begin{aligned}
\bar q_{t,k}
&=\frac{1}{|G_{t,k}|}\sum_{n\in G_{t,k}}
\sigma\!\left(
\frac{S(m_t,e_{t,n})-S(m_t,e_t^{\mathrm{base}})}
{T_{\mathrm{cal}}}\right),\\
A_{t,k}&=\operatorname{sgn}(\bar q_{t,k}-\tfrac12)
\bigl[|\bar q_{t,k}-\tfrac12|-\epsilon_{\mathrm{tie}}\bigr]_+.
\end{aligned}
\end{equation}
Here, $[x]_+=\max(x,0)$, and $\bar q_{t,k}$ is a bounded base-relative group score, not an outcome probability.
The sign of $A_{t,k}$ marks above- or below-base preference, and its magnitude retains only excess beyond the tie band; all group members share $A_{t,k}$.
The fixed blank-instruction effect $e_t^{\mathrm{base}}$ avoids declaring a winner when every candidate scores below no guidance.
$T_{\mathrm{cal}}$ rescales gaps and $\epsilon_{\mathrm{tie}}$ defines the tie band; both are fixed on matched validation pairs~\cite{guo2017calibration}.
Appendix~\ref{sec:appendix_impl} reports these calibration settings and the selected grouping threshold $\epsilon_g$.

\paragraph{Policy update.}
Let $\mathcal J_{\mathrm{PPO}}(z_{t,n};A_{t,k})$ denote the standard token-averaged clipped PPO objective for candidate $n\in G_{t,k}$, including KL regularization to the fixed supervised policy $\pi_{\mathrm{ref}}$~\cite{schulman2017ppo,shao2024deepseekmath}.
Let $K_t^+$ be the number of groups with $A_{t,k}\neq0$; if $K_t^+=0$, context $t$ contributes no gradient.
Otherwise,
\begin{equation}
\label{eq:egpo_loss}
\mathcal L_{\mathrm{EGPO}}
=-\mathbb E_t\!\left[
\frac{1}{K_t^+}
\sum_{k:A_{t,k}\neq0}
\frac{1}{|G_{t,k}|}\sum_{n\in G_{t,k}}
\mathcal J_{\mathrm{PPO}}(z_{t,n};A_{t,k})
\right].
\end{equation}
Here, $\mathbb E_t$ averages eligible contexts; the nested group and candidate averages give each active group equal total weight and divide it among its members, while token averaging remains inside $\mathcal J_{\mathrm{PPO}}$.
Only the instruction policy is updated; S-DT, CEV, and $\pi_{\mathrm{ref}}$ remain frozen.
Deployment uses only the instruction policy and S-DT.

\section{Experiments}
\label{sec:experiments}

\subsection{Experimental Setup}
\label{sec:exp_setup}
\begin{table*}[!t]
\centering
\caption{Overall Score ($\uparrow$) under five budget ratios. Best and second-best results are bold and underlined.}
\label{tab:overall_performance}
\small
\setlength{\tabcolsep}{4pt}
\renewcommand{\arraystretch}{0.95}
\resizebox{\textwidth}{!}{%
\begin{tabular}{@{}llcccccccccccc@{}}
\toprule
Dataset & Budget & USCB & CQL & IQL & DT & DiffBid & GAVE & GRAD & QGA & JD-BP & LBM & SemBid & LangBP \\
\midrule
\multirow{5}{*}{AuctionNet}
& 50\%  & 86  & 103 & 165 & 185 & 68  & 201 & 204 & 206 & \underline{213} & 192 & 206 & \textbf{223} \\
& 75\%  & 135 & 133 & 233 & 254 & 122 & 296 & 293 & 282 & \underline{321} & 279 & 277 & \textbf{342} \\
& 100\% & 157 & 169 & 284 & 324 & 181 & 376 & 372 & 353 & \underline{398} & 348 & 336 & \textbf{419} \\
& 125\% & 220 & 199 & 343 & 380 & 222 & 421 & 432 & 409 & \underline{451} & 381 & 401 & \textbf{470} \\
& 150\% & 281 & 236 & 389 & 429 & 277 & 467 & 476 & 463 & \underline{478} & 426 & 450 & \textbf{501} \\
\midrule
\multirow{5}{*}{AuctionNet-Sparse}
& 50\%  & 11.5 & 15.4 & 17.5 & 17.6 & 10.6 & 19.6 & 20.0 & 19.6 & \underline{21.2} & 18.7 & 17.6 & \textbf{21.8} \\
& 75\%  & 14.9 & 21.3 & 25.5 & 24.4 & 21.0 & 28.3 & 28.5 & 29.0 & \underline{32.9} & 26.7 & 24.6 & \textbf{34.0} \\
& 100\% & 17.5 & 26.8 & 30.5 & 28.5 & 23.8 & 37.2 & 37.4 & 38.8 & \underline{42.5} & 34.3 & 33.1 & \textbf{43.4} \\
& 125\% & 26.7 & 31.4 & 34.0 & 35.2 & 30.6 & 42.7 & 43.2 & \underline{45.3} & 44.9 & 38.6 & 37.6 & \textbf{46.2} \\
& 150\% & 31.3 & 37.2 & 40.0 & 40.5 & 35.5 & 47.4 & 47.5 & \underline{50.1} & 47.5 & 43.2 & 46.9 & \textbf{51.1} \\
\bottomrule
\end{tabular}%
}
\end{table*}

\paragraph{Datasets.}
AuctionNet variants share rules and 48-step trajectories but differ in conversion density and CPA regime~\cite{su2024auctionnet,zhang2026qga}.

\paragraph{Joint trajectory construction.}
AuctionNet provides bidding trajectories but no post-auction pricing corrections.
For joint-policy training, we augment the released multi-agent simulator with a PID pricing-correction controller and rerun delivery on the original exogenous traffic~\cite{su2024auctionnet,meng2026jdbp}.
At each interval, actions are determined from pre-decision campaign states, and pricing corrections are applied only to winning impressions after GSP settlement.
Corrected charges govern budget feasibility and subsequent state transitions, thereby affecting later eligibility and competition, whereas original auction charges are retained for KPI accounting.
These closed-loop rollouts are used only to construct training data; the PID controller is absent during evaluation.
The data-construction code and constructed training data will be released upon publication.

\paragraph{Evaluation protocol.}
We use the released AuctionNet offline replay evaluator~\cite{su2024auctionnet,meng2026jdbp,zhang2026qga} over all 48 advertisers at five budget ratios, $\{0.5,0.75,1.0,1.25,1.5\}$.
For LangBP, pricing corrections are applied after replayed wins and update budget feasibility and the policy state, with logged market thresholds fixed across methods.

\paragraph{Baselines.}
Baselines include RL methods \textbf{USCB}, \textbf{CQL}, and \textbf{IQL}~\cite{he2021uscb,kumar2020cql,kostrikov2022iql}; generative methods \textbf{DT}, \textbf{DiffBid}, \textbf{GAVE}, \textbf{GRAD}, and \textbf{QGA}~\cite{chen2021decision,guo2024diffbid,gao2025gave,lei2026grad,zhang2026qga}; joint-control \textbf{JD-BP}~\cite{meng2026jdbp}; and language-guided \textbf{LBM} and \textbf{SemBid}~\cite{li2026lbm,zhu2026sembid}.
We use full variants; LBM denotes LBM(GQPO), and bidding-only methods set $y_i=0$.
Appendix~\ref{sec:appendix_exp} reports \textbf{GUIDE}~\cite{zhang2026guide} and \textsc{LangBP w/o Pricing} for action-space isolation.

\paragraph{Metrics.}
Following recent AuctionNet-based studies~\cite{meng2026jdbp,zhang2026qga}, the primary metric is
\begin{equation}
\label{eq:auctionnet_score}
\mathrm{Score}
=
V\cdot
\min\left(
\frac{\mathrm{tCPA}\,V}{C_{\mathrm{ori}}},
1
\right)^2,
\end{equation}
where $V=\sum_i x_i v_i$ is conversion value and $C_{\mathrm{ori}}=\sum_i x_i c_i$ is the original auction cost.
Consistent with Eq.~\eqref{eq:jom}, corrected charges govern actual expenditure and future budgets, while the CPA term retains original auction costs for bidding-only comparability.

\paragraph{Implementation details.}
We use a frozen GPT-5.5 teacher and a deployable Qwen-3.5-0.8B instruction policy.
Results average five independent runs, with validation data used for hyperparameter and checkpoint selection.
Appendices~\ref{sec:appendix_impl}--\ref{sec:appendix_components} report training configurations, loss weights, validation-selected thresholds, Relative-Q settings, five-run statistics, and diagnostics; code will be released.

\subsection{Overall Performance (RQ1)}
\label{sec:overall}

\paragraph{RQ1.}
\textit{Does LangBP consistently outperform strong baselines across the two benchmark variants and different budget ratios?}

Table~\ref{tab:overall_performance} reports the Score on AuctionNet and AuctionNet-Sparse under five budget ratios.
LangBP achieves the highest Score in all ten evaluated settings among the compared methods.
On AuctionNet, JD-BP is the strongest baseline at all five budget ratios, while LangBP improves over it by 4.2\%--6.5\%.

On AuctionNet-Sparse, the strongest baseline varies with the budget ratio: JD-BP performs best among the baselines at 50\%--100\%, whereas QGA performs best at 125\% and 150\%.
LangBP exceeds the corresponding strongest baseline by approximately 2.0\%--3.3\%.

Among the language-guided methods, LangBP also achieves higher Scores than LBM and SemBid in every evaluated setting.

\subsection{Ablation Study (RQ2)}
\label{sec:ablation}

\paragraph{RQ2.}
\textit{How do LangBP's design choices affect performance?}

All ablations use AuctionNet-Sparse at five budget ratios and follow the same five-run protocol as Section~\ref{sec:exp_setup}.

We consider five controlled variants.
\textit{SFT Only} uses the supervised instruction policy before post-training with execution feedback and retains the corresponding frozen S-DT executor.
\textit{Direct Action} replaces inverse action recovery from the predicted target state with a joint-action head that directly consumes the instruction-conditioned representation, while retaining target-state prediction and the downstream CEV--EGPO pipeline.
\textit{Relative-Q Feedback} replaces CEV with a frozen offline action-value critic and computes candidate feedback relative to the action induced by the same blank instruction, while retaining support filtering, effect grouping, and the downstream policy-update procedure.
\textit{w/o Explanation} replaces the free-form explanation with a fixed blank string while retaining the bidding-direction and pricing-strength fields and the rest of the pipeline.
Finally, \textit{w/o Effect Grouping} treats each supported candidate as a singleton group, reducing effect-grouped credit assignment to candidate-wise optimization.

\begin{table}[t]
\centering
\caption{AuctionNet-Sparse ablations (five-run means). Best and second-best results are bold and underlined.}
\label{tab:ablation}
\setlength{\tabcolsep}{2pt}
\renewcommand{\arraystretch}{1.02}
\begin{tabular*}{\columnwidth}{@{\extracolsep{\fill}}lccccc@{}}
\toprule
Variant & 50\% & 75\% & 100\% & 125\% & 150\% \\
\midrule
SFT Only
& 20.0 & 30.6 & 39.9 & 42.7 & 46.8 \\
Direct Action
& 20.7 & 31.9 & 41.1 & 44.4 & 48.8 \\
Relative-Q Feedback
& 21.2 & 32.3 & 41.7 & 44.9 & 49.4 \\
w/o Explanation
& \underline{21.3} & \underline{33.2} & 42.2 & 45.1 & 49.9 \\
w/o Effect Grouping
& 21.0 & 33.1 & \underline{42.8}
& \underline{45.4} & \underline{50.2} \\
\textbf{LangBP}
& \textbf{21.8} & \textbf{34.0} & \textbf{43.4}
& \textbf{46.2} & \textbf{51.1} \\
\bottomrule
\end{tabular*}
\end{table}

Post-training with execution feedback improves over the supervised initialization by 8.2\%--11.1\% across the five budget ratios, the largest gap among the controlled variants.
LangBP's Score exceeds \textit{Direct Action} by 4.1\%--6.6\% across five settings, supporting target-state mediation over direct action prediction.

LangBP improves over \textit{Relative-Q Feedback} by 2.8\%--5.3\% under the same downstream optimization framework.
Blanking the free-form explanation while retaining the two categorical control fields lowers Score by 2.3\%--2.8\%.
The gain over \textit{w/o Effect Grouping} is smaller (1.4\%--3.8\%) but remains positive at all five budget ratios.

\subsection{Component Analysis (RQ3)}
\label{sec:component_analysis}

\paragraph{RQ3.}
\textit{Do S-DT, CEV, and EGPO exhibit their intended execution, ranking, and credit-allocation behavior?}

We complement the performance ablations with split-specific diagnostics on AuctionNet-Sparse.

\begin{figure}[t]
\centering
\includegraphics[width=\columnwidth]{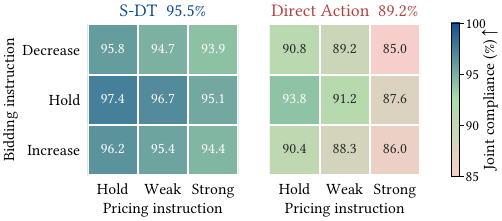}
\Description{Two shared-scale heatmaps compare joint instruction compliance for S-DT and Direct Action across nine bidding and pricing instruction combinations.}
\caption{Joint instruction compliance across the $3\times3$ bidding--pricing instruction combinations on AuctionNet-Sparse.}
\label{fig:component_sdt}
\end{figure}

\paragraph{S-DT execution.}
We issue all $3\times3$ bidding--pricing instructions to S-DT and the Direct Action variant from Section~\ref{sec:ablation} on the same validation contexts.
Joint compliance requires both executed controls to satisfy their instruction-specified categories.
Figure~\ref{fig:component_sdt} reports macro joint compliance of $95.5\%$ for S-DT and $89.2\%$ for Direct Action, with worst-cell compliance of $93.9\%$ and $85.0\%$, respectively.
S-DT is more compliant across all nine combinations.

\begin{figure}[t]
\centering
\includegraphics[width=\columnwidth]{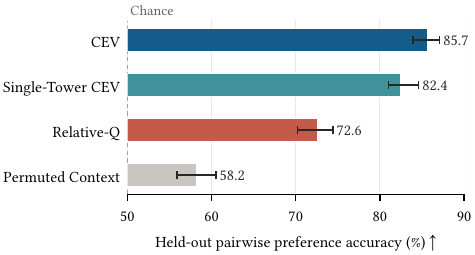}
\Description{Four horizontal bars with confidence intervals compare validation pairwise preference accuracy for CEV, a Single-Tower CEV, Relative-Q, and CEV under a supported context permutation; a dashed vertical line marks chance accuracy.}
\caption{Validation preference accuracy on AuctionNet-Sparse using the same behavior-supported pairs. Error bars are advertiser-cluster bootstrap 95\% confidence intervals.}
\label{fig:component_cev}
\end{figure}

\paragraph{CEV ranking.}
Residualized suffix utility is used only to construct pairwise preference labels and is not provided as input to any scorer.
Figure~\ref{fig:component_cev} shows pairwise accuracy of $85.7\%$ for CEV.
The capacity-matched Single-Tower CEV concatenates $m$ and $e$ at input and processes them with one MLP, removing the separate encoders and explicit embedding-product interaction; it reaches $82.4\%$.
The gap suggests that factorized context--effect encoding is useful for this ranking task.
Relative-Q ranks the same pairs with a frozen twin-$Q$ critic and reaches $72.6\%$, consistent with the dedicated ranking objective being better aligned with the execution-feedback signal.
Accuracy falls to $58.2\%$ after support-preserving context permutation, supporting the importance of correct context--effect correspondence.
These results concern observational ranking within behavior-supported regions, not causal estimation.

\begin{figure}[t]
\centering
\includegraphics[width=\columnwidth]{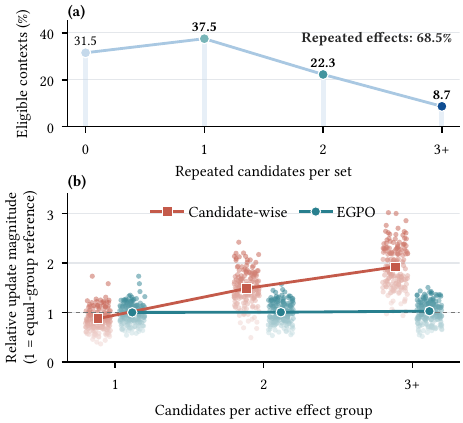}
\Description{The first panel shows the distribution of the number of redundant supported candidates, computed as the supported-candidate count minus the number of distinct effect groups. Positive values account for 68.5 percent of eligible contexts. The second shows equal-size subsamples of group-level update magnitudes and their full-sample medians under candidate-wise weighting and EGPO at three group multiplicities.}
\caption{EGPO allocation diagnostics: (a) redundant supported candidates after effect grouping, where positive values indicate repeated effects; and (b) relative update magnitude by active-group multiplicity. Panel (b) shows equal-size subsamples; lines and correlations use all active groups.}
\label{fig:component_egpo}
\end{figure}

\paragraph{EGPO allocation.}
On held-out evaluation candidate sets from all 48 advertisers, Figure~\ref{fig:component_egpo}(a) plots, for each eligible context, the number of redundant supported candidates after effect grouping, computed as the supported-candidate count minus the number of distinct effect groups.
Its positive bins contain $68.5\%$ of eligible candidate sets, indicating repeated effects, while $K_t$ averages $5.27$.
To isolate outer credit allocation, we hold the candidate sets, effect groups, CEV scores, and group advantages fixed and recompute the absolute clipped-PPO update coefficient, normalized within each context by the equal-group reference.
The candidate-wise magnitude rises from a median of $0.88$ for singleton groups to $1.92$ for groups with at least three candidates, with a Spearman rank correlation of $0.56$.
Figure~\ref{fig:component_egpo}(b) shows that EGPO keeps the corresponding medians between $1.00$ and $1.03$, leaving only a near-zero Spearman rank correlation with multiplicity (approximately $0.00$).
This diagnostic isolates multiplicity-dependent allocation; Table~\ref{tab:ablation} remains the controlled evidence for downstream performance.

\subsection{Online A/B Test}
\label{sec:online}
We further evaluate LangBP with a seven-day online A/B test on a large-scale e-commerce advertising platform, using a strong production policy from the joint bidding-and-pricing paradigm~\cite{meng2026jdbp} as the baseline.
The test runs on the target-ROI (tROI) product, an instance of the ROI-constrained problem in Eq.~\eqref{eq:jom}: advertisers set a budget and a target ROI, and the system bids and prices to meet it.
Its traffic is sufficient for statistically reliable measurement.
In deployment, both levels of LangBP run off the real-time auction path.
The $0.8$B student instruction policy refreshes each campaign's high-level strategy once per hour, completing a full pass over all active campaigns well within each cycle.
The lighter S-DT executor then updates the bidding and pricing actions every ten minutes from the current instruction and the latest campaign state.

Four platform-level metrics are reported.
\emph{Ad Click} and \emph{Ad Cost} are the number of ad clicks and the total ad spend.
\emph{Ad Value} is the ad revenue converted from advertising GMV, which reflects the value delivered to advertisers.
\emph{Achievement} is the fraction of campaigns whose realized ROI falls within $[0.8,1.2]$ of the target.
As shown in Table~\ref{tab:online}, LangBP improves Ad Click by $2.59\%$, Ad Cost by $1.86\%$, Ad Value by $3.34\%$, and Achievement by $1.83$ percentage points over the production policy.
Ad Value grows more than Ad Cost, so each unit of spend delivers more advertiser value.
Together with the higher Achievement, this shows that the revenue gain comes from more efficient value delivery and better target attainment, not from a relaxed ROI target.

{\setlength{\intextsep}{4pt}
\begin{table}[H]
\centering
\captionsetup{skip=3pt}
\caption{Online A/B test on the tROI bidding product.
Each value is the relative gain of the LangBP bucket over the production bucket;
Achievement is in percentage points (pp).}
\label{tab:online}
\setlength{\tabcolsep}{4pt}
\renewcommand{\arraystretch}{1.02}
\begin{tabular*}{\columnwidth}{@{\extracolsep{\fill}}cccc@{}}
\toprule
Ad Click & Ad Cost & Ad Value & Achievement \\
\midrule
$+2.59\%$ & $+1.86\%$ & $+3.34\%$ & $+1.83$\,pp \\
\bottomrule
\end{tabular*}
\end{table}
}

\section{Conclusion}
\label{sec:conclusion}
We present LangBP, a hierarchical framework for language-guided joint bidding and pricing.
S-DT maps instructions to explicit target states and recovers joint actions through inverse dynamics.
CEV provides context-dependent observational preferences over behavior-supported execution effects, while EGPO balances policy updates across effect groups.
Across AuctionNet and AuctionNet-Sparse, LangBP achieves higher Scores than the evaluated numerical, generative, and language-guided baselines, while ablation and component analyses provide complementary evidence for its design choices and their intended behavior.
The online A/B test further reports gains in ad value and target achievement on a large-scale advertising platform.
CEV is limited to observational evidence in behavior-supported regions; future work could study online feedback under distribution shift.

\appendix
\raggedbottom

\section{Additional Results}
\label{sec:appendix_exp}

\paragraph{Action-space-controlled comparison.}
Table~\ref{tab:without_pricing} compares full LangBP with bidding-only GUIDE, LBM, SemBid, and LangBP w/o Pricing on AuctionNet-Sparse~\cite{zhang2026guide,li2026lbm,zhu2026sembid}.
The last sets $a_t^{\mathrm{price}}=0$; all else, including the data split and replay evaluation, is unchanged.

{\setlength{\intextsep}{4pt}
\begin{table}[H]
\centering
\captionsetup{skip=2pt}
\caption{Action-space comparison on AuctionNet-Sparse.}
\label{tab:without_pricing}
\footnotesize
\setlength{\tabcolsep}{1.6pt}
\renewcommand{\arraystretch}{0.90}
\begin{tabular*}{\columnwidth}{@{\extracolsep{\fill}}llccccc@{}}
\toprule
Method & Action & 50\% & 75\% & 100\% & 125\% & 150\% \\
\midrule
GUIDE & Bidding & 20.3 & 29.1 & 37.6 & 43.3 & 48.3 \\
LBM & Bidding & 18.7 & 26.7 & 34.3 & 38.6 & 43.2 \\
SemBid & Bidding & 17.6 & 24.6 & 33.1 & 37.6 & 46.9 \\
LangBP w/o Pricing
& Bidding & \underline{20.8} & \underline{31.7} & \underline{40.8}
& \underline{44.7} & \underline{49.5} \\
\textbf{LangBP}
& Joint & \textbf{21.8} & \textbf{34.0} & \textbf{43.4}
& \textbf{46.2} & \textbf{51.1} \\
\bottomrule
\end{tabular*}
\end{table}
}

\section{Implementation Details}
\label{sec:appendix_impl}

Validation-selected hyperparameters and checkpoints are fixed before final evaluation.

\paragraph{Data and models}
We use AuctionNet and AuctionNet-Sparse, with GPT-5.5 as the teacher and Qwen-3.5-0.8B as the instruction policy.

\paragraph{SFT and S-DT}
Instruction labels use the 33rd percentile of nonzero $|\Delta a^{\mathrm{bid}}|$ and the 33rd/67th percentiles of $|a^{\mathrm{price}}|$.
SFT uses AdamW, learning rate $10^{-5}$, batch size 64, and five epochs.
S-DT uses history length 20, four blocks per stream, hidden size 128, four attention heads, FFN size 512, and eight queries; numerical pretraining and alignment use learning rates $10^{-4}$ and $5{\times}10^{-5}$.
The dimension-averaged coefficients (state, action, instruction, dynamics) are $(1,.5,.5,.1)$ for observed samples, $(0,0,.5,.1)$ for counterfactual samples, and $(.25,.125,0,.025)$ for blank-instruction samples, with $\alpha=1$.

\paragraph{CEV and EGPO}
CEV uses suffix horizon 12 and five advertiser folds, fitting each matching-vector residualizer on four folds and predicting the held-out fold; its learning rate is $3{\times}10^{-4}$ with batch size 512.
EGPO samples eight candidates with temperature 0.9 and top-$p$ 0.98; policy optimization uses learning rate $10^{-6}$, PPO clip 0.2, and KL coefficient 0.03.
Experiments use two NVIDIA B200 GPUs and five independent runs.

Table~\ref{tab:appendix_selection} uses $K_a$ for the neighborhood size in counterfactual-action support checking and $K$ for the CEV effect-neighborhood size in Eq.~\eqref{eq:cev_support}; the support cutoff is the upper bound on $D_{\mathrm{sup}}$.
For matched-pair construction, $\epsilon_c$ scales context-similarity weights, $\epsilon_e$ is the minimum normalized RMS separation between two effects, and $\epsilon_u$ is the minimum absolute residual-utility gap.
The remaining quantities retain their roles from Section~\ref{sec:egpo}: $T_{\mathrm{cal}}$ rescales CEV score gaps, $\epsilon_{\mathrm{tie}}$ defines the blank-relative tie band, and $\epsilon_g$ bounds within-group effect distance.

\begin{table}[H]
\centering
\caption{Validation-selected support, ranking, and grouping settings.}
\label{tab:appendix_selection}
\footnotesize
\setlength{\tabcolsep}{2.5pt}
\begin{tabular}{@{}lp{0.39\columnwidth}cp{0.18\columnwidth}@{}}
\toprule
Parameter & Candidates or rule & Selected & Split \\
\midrule
$K_a$ & $\{10,20,40\}$ & 20 & Validation \\
$K$ & $\{16,32,64\}$ & 32 & Validation \\
$\epsilon_c$ & $\{0.6,0.8,1.0\}$ & 0.8 & Validation \\
Support cutoff & $\{0.3,0.4,0.5\}$ & 0.4 & Validation \\
$\epsilon_e$ & $\{0.15,0.20,0.25\}$ & 0.20 & Validation \\
$\epsilon_u$ & positive utility-gap quantiles $\{50,60,70\}$ & 60th & Validation \\
$T_{\mathrm{cal}}$ & logistic calibration & 0.41 & Validation \\
$\epsilon_{\mathrm{tie}}$ & 95th near-tie percentile & 0.06 & Validation \\
$\epsilon_g$ & $\{0.24,0.30,0.36\}$ & 0.30 & Validation \\
\bottomrule
\end{tabular}
\end{table}

\paragraph{Relative-Q Feedback}
The frozen IQL-style critic receives numerical pre-decision context and normalized joint actions, but neither instructions nor predicted target-state changes~\cite{kostrikov2022iql,li2025gas,li2026lbm}.
It uses a shared six-layer causal Transformer encoder (history 20, hidden size 512, eight heads) with twin-$Q$ and value heads.
AdamW uses learning rate $10^{-4}$, weight decay $10^{-2}$, batch size 128, and 400k updates; discount factor 0.99, expectile 0.7, and target update rate 0.01, with one-step TD over 48-step episodes.
The training-reward scale is $s_Q=\operatorname{Quantile}_{0.95}^{\mathrm{train}}(|r_t|)$, and the clipped critic reward is $r_t^Q=\operatorname{clip}(r_t/s_Q,-5,5)$.
For blank-relative candidate feedback, $T_Q=0.30$ rescales critic score gaps and $\epsilon_{\mathrm{tie}}^Q=0.05$ defines the tie band.

\paragraph{Pricing-action execution}
For a winning impression $i$ in interval $t$, the simulator projects the requested additive pricing action $a_t^{\mathrm{price}}\leq0$ to
\begin{equation}
p_i=\min\!\left\{c_i,\max\!\left(c_i+a_t^{\mathrm{price}},(1-r_{\max})c_i,p_{\min}\right)\right\},
\qquad y_i=p_i-c_i,
\end{equation}
where $r_{\max}\in[0,1]$ is the refund-rate cap and $p_{\min}\geq0$ is the minimum charge; losing impressions have $p_i=y_i=0$.
The action therefore cannot change current GSP allocation or increase the auction charge. Its realized correction may vary with $c_i$ because of these execution constraints.

\paragraph{Preprocessing and support checks}
Execution-effect components are normalized with training statistics and clipped to $[-5,5]$.
Counterfactual actions must pass a $K_a$-nearest-neighbor action-support check fixed on validation.
CEV pairs use distinct logged periods from the same advertiser, decision step, and budget configuration.
Retained pairs use context-similarity weight $w_{ij}=\exp[-\operatorname{RMS}(m_i,m_j)/\epsilon_c]$.

\section{Evaluation Details}
\label{sec:appendix_eval}

\subsection{Five-Run Results}

Tables~\ref{tab:appendix_overall_std} and~\ref{tab:appendix_ablation_std} report dispersion across the same five runs used for the means in Tables~\ref{tab:overall_performance} and~\ref{tab:ablation}.

\begin{table}[H]
\centering
\caption{LangBP Score (mean $\pm$ standard deviation).}
\label{tab:appendix_overall_std}
\footnotesize
\setlength{\tabcolsep}{3pt}
\begin{tabular}{@{}lccccc@{}}
\toprule
Dataset & 50\% & 75\% & 100\% & 125\% & 150\% \\
\midrule
AuctionNet & $223\!\pm\!1.8$ & $342\!\pm\!2.4$ & $419\!\pm\!2.1$ & $470\!\pm\!2.7$ & $501\!\pm\!2.5$ \\
AuctionNet-Sparse & $21.8\!\pm\!0.14$ & $34.0\!\pm\!0.21$ & $43.4\!\pm\!0.19$ & $46.2\!\pm\!0.26$ & $51.1\!\pm\!0.24$ \\
\bottomrule
\end{tabular}
\end{table}

\begin{table}[H]
\centering
\caption{AuctionNet-Sparse ablations (mean $\pm$ standard deviation).}
\label{tab:appendix_ablation_std}
\scriptsize
\setlength{\tabcolsep}{1.1pt}
\begin{tabular}{@{}lccccc@{}}
\toprule
Variant & 50\% & 75\% & 100\% & 125\% & 150\% \\
\midrule
SFT Only & $20.0\!\pm\!0.24$ & $30.6\!\pm\!0.31$ & $39.9\!\pm\!0.38$ & $42.7\!\pm\!0.34$ & $46.8\!\pm\!0.42$ \\
Direct Action & $20.7\!\pm\!0.20$ & $31.9\!\pm\!0.27$ & $41.1\!\pm\!0.31$ & $44.4\!\pm\!0.29$ & $48.8\!\pm\!0.35$ \\
Relative-Q Feedback & $21.2\!\pm\!0.18$ & $32.3\!\pm\!0.25$ & $41.7\!\pm\!0.28$ & $44.9\!\pm\!0.26$ & $49.4\!\pm\!0.32$ \\
w/o Explanation & $21.3\!\pm\!0.18$ & $33.2\!\pm\!0.24$ & $42.2\!\pm\!0.27$ & $45.1\!\pm\!0.28$ & $49.9\!\pm\!0.31$ \\
w/o Effect Grouping & $21.0\!\pm\!0.16$ & $33.1\!\pm\!0.22$ & $42.8\!\pm\!0.24$ & $45.4\!\pm\!0.25$ & $50.2\!\pm\!0.29$ \\
LangBP & $21.8\!\pm\!0.14$ & $34.0\!\pm\!0.21$ & $43.4\!\pm\!0.19$ & $46.2\!\pm\!0.26$ & $51.1\!\pm\!0.24$ \\
\bottomrule
\end{tabular}
\end{table}

Matched seeds give LangBP-minus-w/o-Effect-Grouping Score differences of $0.80\!\pm\!0.17$, $0.90\!\pm\!0.22$, $0.60\!\pm\!0.14$, $0.80\!\pm\!0.20$, and $0.90\!\pm\!0.19$ from 50\% to 150\% budget; the difference is positive for all five seeds at every budget ratio.

\section{Component-Analysis Details}
\label{sec:appendix_components}

\subsection{CEV Diagnostics}

All four comparisons use the same 4,231 validation pairs from the 10 advertisers excluded from scorer fitting.
Resampling these advertisers as clusters 2,000 times gives overall 95\% confidence intervals of $[83.9,87.1]$, $[81.0,84.6]$, $[70.2,74.4]$, and $[55.9,60.5]$ percent in the order shown in Figure~\ref{fig:component_cev}; support filtering retains $81.6\%$ of sampled candidates.
The support-preserving permutation fixes the two effects and their preference label, substitutes a context from another validation advertiser at the same decision step and budget configuration, and rechecks support for both effects.
The Single-Tower CEV concatenates context and effect before one capacity-matched MLP and is retrained on the same training pairs.

Support regions are training-set tertiles, fixed before evaluation, of the worst cross-context support distance among the two effects in a matched pair:
\begin{equation}
\max_{\substack{\chi\in\{\chi_i,\chi_j\}\\e\in\{e_i,e_j\}}}
D_{\mathrm{sup}}(e\mid t,\chi).
\end{equation}
Table~\ref{tab:appendix_cev} gives the corresponding pair counts and accuracies.

\begin{table}[H]
\centering
\caption{Held-out pairwise preference accuracy (\%) across support regions.}
\label{tab:appendix_cev}
\footnotesize
\setlength{\tabcolsep}{4pt}
\begin{tabular}{@{}lccc@{}}
\toprule
 & High & Medium & Near boundary \\
\midrule
Pairs & 1,384 & 1,407 & 1,440 \\
CEV & 90.8 & 86.4 & 80.0 \\
Single-Tower CEV & 88.1 & 82.1 & 77.3 \\
Relative-Q & 78.2 & 73.1 & 66.6 \\
\bottomrule
\end{tabular}
\end{table}

\subsection{EGPO Diagnostics}

At grouping thresholds $\{0.8\epsilon_g,\epsilon_g,1.2\epsilon_g\}$, repeated-effect rates are $65.5\%$, $68.5\%$, and $71.5\%$, with 5.46, 5.27, and 5.05 distinct groups on average.
This sensitivity check uses the same eligible contexts and supported candidates at all three thresholds.

For Figure~\ref{fig:component_egpo}(b), candidate-wise and effect-grouped weighting use the same supported candidates, groups, CEV scores, and nonzero group advantages.
We aggregate the absolute clipped-PPO coefficient on completion-token log probabilities, excluding KL, into group mass $M_{t,k}$ and report
\begin{equation}
R_{t,k}=\frac{K_t^+M_{t,k}}{\sum_jM_{t,j}},
\end{equation}
where $R_{t,k}=1$ denotes equal allocation among the $K_t^+$ active groups.
Here $R_{t,k}$ is the relative update mass of group $k$, $M_{t,k}$ is its aggregated absolute update coefficient, and the denominator sums this mass over active groups.
If $N_t$ denotes the number of supported candidates in context $t$, candidate-wise candidates receive outer weight $1/N_t$, whereas EGPO candidates in group $G_{t,k}$ receive $1/(K_t^+|G_{t,k}|)$.
Contexts with fewer than two active groups or zero total coefficient are excluded.

The analysis contains 5,686 active groups from 1,194 held-out evaluation contexts across all 48 advertisers, covering both singleton and multi-candidate groups.
Figure~\ref{fig:component_egpo}(b) draws an equal-size subsample from each multiplicity bin for visibility; medians and Spearman correlations use all groups.
Candidate-wise medians rise from 0.88 to 1.92 with a Spearman rank correlation of $0.56$, whereas EGPO medians remain between 1.00 and 1.03 with a near-zero Spearman rank correlation of $-0.0023$.

\printbibliography

\end{document}